# Ferromagnetic Resonance in selected nanostructural materials designed for technological applications

Ferromagnetic resonance group , Department of Physics, University of Maryland, 2221 Physics Bldg College Park, MD 20742.

Contributors:  G.V. Kurlyandskaya (galina@uniovi.es), S.M. Bhagat (sb27@umail.umd.edu)

## 1. INTRODUCTION

During the past ten years nanostructured materials have been subject of very active research. Fabrication of such systems follows well developed methods [1-2]. The increase in the number of materials available for research and applications requires that the methods of characterization of nanostructural materials be even more precise then before.

In the widest possible sense nanostructured materials are systems which contain structural elements with at least one dimension of nanometer size. For magnetic films or multilayered structures it can be the thickness of the magnetic layers or even the thickness of the fillers between the  layers. For powders the relevant scale is the calliper size of the particles The present studies are focused on magnetic nanostructured systems.

Thin film or multilayered structures have many advantages for technological applications because of compatibility with integrated circuit design.  The magnetoimpedance, MI (change of impedance of a ferromagnet on application of a field)  in 3-layered structures consisting of two magnetic layers separated by a non-magnetic conductive layer has been predicted to show a much higher value compared with that of single layered films of the same composition at optimized thickness [4]. Although a number of theoretical models were developed for different geometrical parameters and different anisotropies of magnetic components in giant MI-sandwiches, experimental tests of the effects of particular anisotropies, stress distributions, interface quality have not been done even in a rough approximation. In many cases the experimental values of MI effect are much smaller than the theoretical predictions. This can be partially explained by appeal to non-ideal anisotropies, mixed interfaces in case of multilayered structures, or by making general comments about non-uniform, uncontrolled, or inappropriate stress distributions. However, more careful characterization of the samples is a must.

Accordingly, the first part of the present research deals with a ferromagnetic resonance, FMR, study of thin films and multilayers containing $Fe_{20}Ni_{80}$ layered nanocomponents. The Permalloy composition was selected for this study because the preparation technology of FeNi films is very well established, the FMR in single layered films has been studied for decades [5], and some data on the MI in FeNi films and GMI sandwiches can be found in the literature [6].

The second system proposed for FMR study consists of Rare Earth – Transition Metal, RE-TM, amorphous alloys prepared by rf- sputtering. They were intensively studied many years ago as materials which can show uniaxial perpendicular anisotropy and were proposed as promising magnetic memory candidates for information storage via cylindrical bubble domains [7].

Recently RE/TM (RE = Gd, Tb; TM = Fe, Co) systems have again attracted significant interest being in this case multilayered structures of nanoscale thickness of the RE and TM components sometimes with additional thin layers of nonmagnetic conducting, semiconductor or insulator solids (Cu, Si, TiN) [8]. These can be used in magnetic optical recording or spin valve structures besides being simply good magnetic systems to study the exchange interactions between two ferrimagnetically coupled sub-systems [9]. In order to study the exchange interaction between pure component systems (Gd) one has to work at temperatures significantly below 270 K. It is well known that the Curie temperature of bulk Gd is close to room temperature. In a nanoscale thickness layer it is further reduced (close to 250 K for a 150 nm film). Therefore, the Co/GdCo system was chosen as a model system both for convenience and in view of possible applications because the component containing the rare earth is itself magnetic at room temperature.

The third group of magnetic materials for FMR characterization consists of powders: commercial polystyrene beads (Dynabeads-480) which are widely used for magnetic separation and proposed as magnetic markers for biosensors and CoNi powders of three sizes with nanoscale particle dimensions. These particles have many applications which include biomedical devices [10] and as magnetic markers [11].

FMR and microwave absorption in micron size powders have been studied previously [12]. More recently new methods of small particle fabrication like the polyol process have been developed. Metallic particles smaller than micrometer size are widely available now [13-14]. Therefore their characterization by microwave methods is highly desirable. To our knowledge, apart from some size effect measurements in zero field [13] there have been no systematic microwave magneto absorption studies of these powders.

## 2. SAMPLES

### GMI related thin films and multilayers

FeNi thin films and structures containing FeNi and Cu layered elements were prepared by rf-sputtering onto glass substrates using a $Ni_{80}Fe_{20}$ alloy target with a base vacuum of $10^{-6}$ Torr with $10^{-3}$ Torr Ar pressure during deposition. The system was previously calibrated with respect to deposition rate of both FeNi and Cu thin films. Therefore the thickness of Permalloy and copper components was estimated by the deposition time at a known deposition rate. The thickness of the FeNi magnetic layer of 2400 Å as it was deduced from the deposition time was checked by the Rutherford Back Scattering method at The University of Maryland and found to be 3500 Å. Therefore a corresponding correction was done and all data reported below represent such corrected thicknesses.

$Ni_{80}Fe_{20}$ composition films have close to zero magnetostriction and therefore stress induced magnetic anisotropy should be rather small. However, incorporation of Ar in the film cannot be ignored. Due to the well known fact that the film composition can be slightly different from that of the target one of the samples was checked at the University of Maryland using EDAX yielding the analysis: Fe 17.3 at.%; Ni 80.36 at%; Si 2.34 at% (the last due to a Si presence in a glass substrate and impossibility of separation of this signal in case of a thin film ) in reasonable agreement with expectation.

Four configurations of the deposited samples are shown in figure 1. The coercive field, $H_c$, and anisotropy field, $H_a$, were estimated by magneto-optical Kerr effect measurements.

1) (1.753 sample) FeNi (3500 Å) single layer film. $H_c \approx 3.5$ - 4 Oe in easy magnetization direction.

2) (1.752 sample) FeNi(1750 Å)/Cu(3500 Å)/FeNi(1750 Å) were deposited as a classic multilayer, i.e. with the central Cu layer having the same geometry as two magnetic layers. $H_c \leq 1$ Oe, $H_a \approx 4$ Oe, prepared without opening the deposition chamber.

3) (1.754 sample) FeNi(1750 Å)/Cu(3500Å)/FeNi(1750 Å) on one half of the substrate and FeNi(1750Å)/FeNi(1750 Å) on the other. In this case the chamber was opened between the deposition of the different layers.

4) FeNi(1750 Å)(Cu(3500 Å))FeNi(1750 Å) magnetoimpedance geometry sándwich with Cu layer of 1.5 mm width which is much narrower than the FeNi layers.

Apart from these thin film samples prepared at Ural State University a thin 500 Å Permalloy film of $Ni_{80}Fe_{20}$ composition was deposited onto glass substrate at Rowan University and kindly made available to us by Prof. S. E. Lofland). The static magnetization of 1753 ($4\pi M_s \approx 10.1$ kOe $M_s \approx 800$ G) determined by SQUID (Fig.2) is in reasonable agreement with the accepted value for $Fe_{20}Ni_{80}$ Permalloy.

Figure 1. $Ni_{80}Fe_{20}$ based thin film structures studied by FMR.

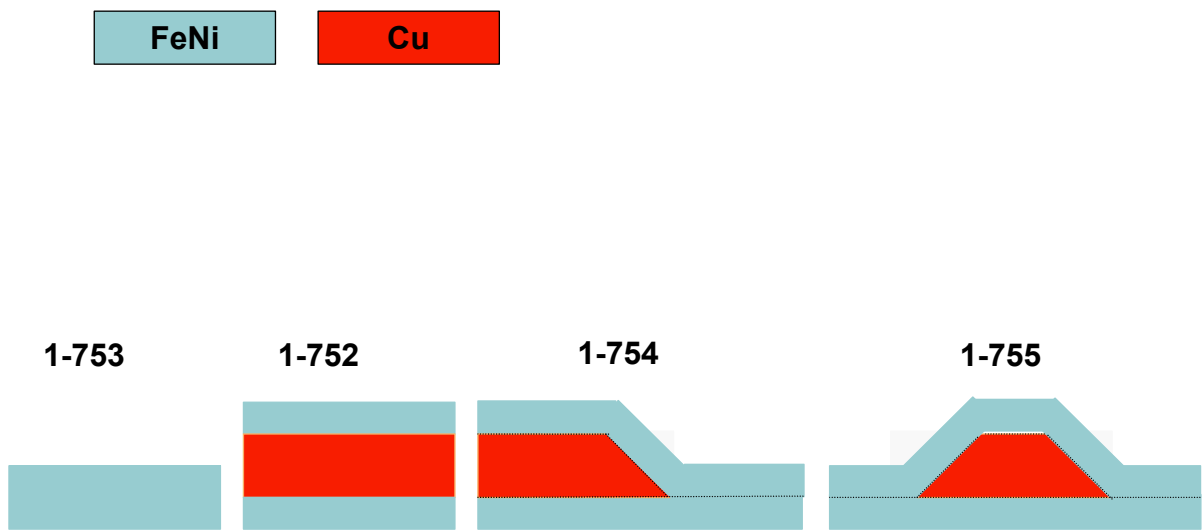

Figure 2: Hystereis loop of Permalloy 3500 Å thin film of 0.31 x 0.495 cm in plane dimensions measured by SQUID: $M_s \approx 800$ Oe (measurements done in collaboration with Dr. Sanjay R. Shinde).

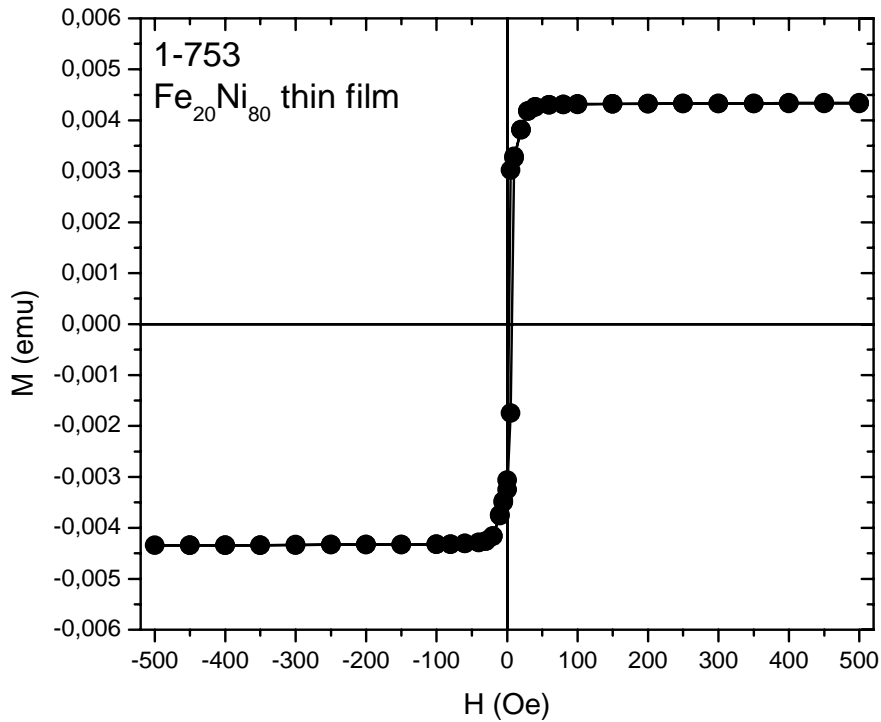

**RE/TM multilayers with non-magnetic sub-layers**

Multilayered structures containing Co and Gd-Co nanoscale layers were prepared by rf-sputtering onto glass substrates using different targets without opening the chamber during depositions. Base vacuum of $10^{-6}$ and $10^{-3}$ Torr Ar pressure were kept constant during corresponding preparation stage. The system was previously calibrated with respect to deposition rate for all the components. Therefore the thickness of the components was estimated by the time of deposition.

Figure 3 shows schematic diagrams of the multilayered structures used in this study and Table 1 gives more data on properties of these structures and their magnetic properties measured by vibrating sample magnetometer, magnetooptical Kerr effect, and rotational anisometer.

Figure 3. Co/ Gd$_{36}$Co$_{64}$ based multilayered structures for FMR study (a) and some characteristic fields definition (b).

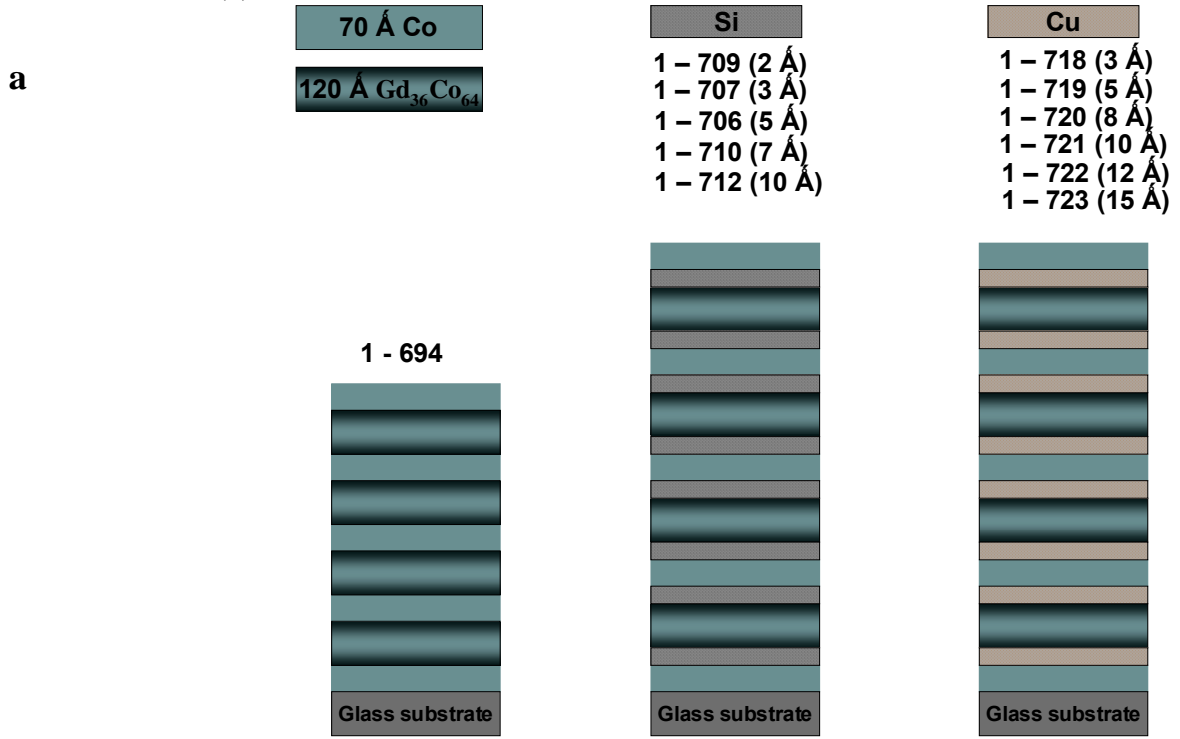

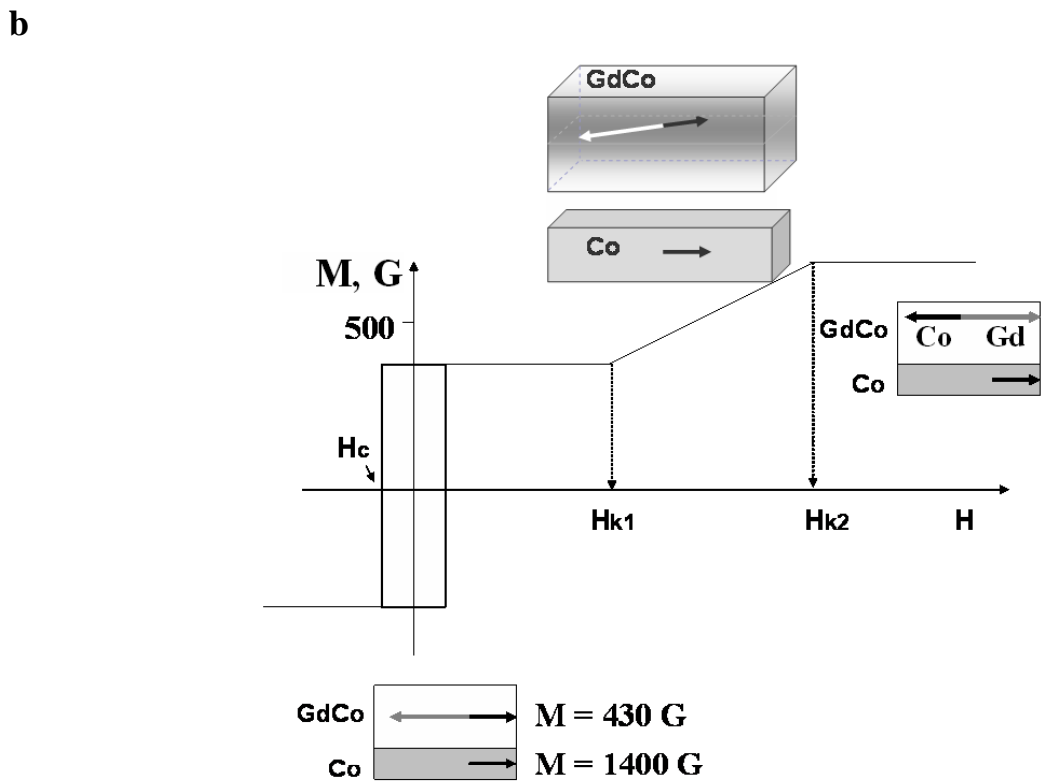

Table 1. Selected properties of multilayered structures and single layered films used for FMR study. $H_c^M$ is a coercive field defined from the MOKE measurements; $H_c^V$ is a coercive field defined from the VSM measurements; $H_{k1}$ is a critical field (figure 3b); $T_{comp}$ (for H = 200 Oe).

| Sample | Structure (Co or Gd is a sign of Kerr effect derived magnetization) | $H_c^M$ (Oe) | $H_c^V$ (Oe) | $H_{k1}$ (Oe) | $T_{comp}$ (K) |
|---|---|---|---|---|---|
| 1.694 | {Co(70 Å) /(Gd-Co)(120 Å)}$_4$/Co(70 Å), **Co** | 16 | 16 | 8000 | < 81 |
| 1.709 | {Co(70 Å)/Si(2 Å)/(Gd-Co)(120 Å) )/Si(2 Å)}$_4$ /Co(70 Å), **Co** | 15-18 | 15 | 3750 | < 81 |
| 1.707 | {Co(70 Å)/Si(3 Å)/(Gd-Co)(120 Å) )/Si(3 Å)}$_4$ /Co(70 Å) , **Co** | 17-21 | 16 | 1000 | < 81 |
| 1.706 | {Co(70 Å)/Si(5 Å)/(Gd-Co)(120 Å) )/Si(5 Å)}$_4$ /Co(70 Å) , **Co** | 13-15 | 9 | 480 | < 81 |
| 1.710 | {Co(70 Å)/Si(7 Å)/(Gd-Co)(120 Å) )/Si(7 Å)}$_4$ /Co(70 Å ), **Co** | 18-25 | 22 | 0 | < 81 |
| 1.712 | {Co(70 Å)/Si(10 Å)/(Gd-Co)(120 Å) )/Si(10 Å)}$_4$ /Co(70 Å) ), **Gd** **Co** | 6 18-20 | 5 15 | 0 | _ |
| 1.718 | {Co(70 Å)/Cu(3 Å)/(Gd-Co)(120 Å) )/Cu(3 Å)}$_4$ /Co(70 Å) | 11-12 | 10 | 8000 | < 81 |
| 1.719 | {Co(70 Å)/Cu(5 Å)/(Gd-Co)(120 Å) )/Cu(5 Å)}$_4$ /Co(70 Å) | 9-15 | 10 | 3300 | < 81 |
| 1.720 | {Co(70 Å)/Cu(8 Å)/(Gd-Co)(120 Å) )/Cu(8 Å)}$_4$ /Co(70 Å) | 5-9 | 5 | 670 | < 81 |
| 1.721 | {Co(70 Å)/Cu(10 Å)/(Gd-Co)(120 Å) )/Cu(10 Å)}$_4$ /Co(70 Å), **Co** **Co** | 8-10 15-30 | 8 15 | 40 | _ |
| 1.722 | {Co(70 Å)/Cu(12 Å)/(Gd-Co)(120 Å) )/Cu(12 Å)}$_4$ /Co(70 Å), **Co** **Co** | 8 15-20 | 9 9 | 0 | _ |
| 1.723 | {Co(70 Å)/Cu(15 Å)/(Gd-Co)(120 Å) )/Cu(15 Å)}$_4$ /Co(70 Å), **Co** **Co** | 12 15-20 | 10 12 | 0 | _ |
| 1199 | Single layered Co film of 300 Å | - | - | - | - |
| 1.703 | Gd$_{36}$Co$_{64}$ single layered film | - | - | - | - |

The saturation magnetization of GdCo single layer film was measured by VSM at room temperature being close to 430 G. It is wellknown that Gd and Co magnetic moments in GdCo system align in opposite directions. For the composition under consideration ($Gd_{36}Co_{64}$) magnetic moment of Gd is higher than that one of the Co at any temperature and therefore there is no compensation state (or by the data of some authors it is very close to the Curie point being accordingly 450 and 500 K [7]) when total gadolinium moment is equal to the moment of Co. Therefore the magnetization direction of the total magnetization of GdCo alloy layer is generally aligned in a direction of the Gd magnetization and - $M_{Gd}$ + $M_{Co}$ = - 430 G (Fig. 3b). In this case for {Co(70 Å) /(Gd-Co)(120 Å)}$_4$/Co(70 Å) structure the total magnetic moment of GdCo layers of area S for room temperature can be written as 4 x 120 Å x (- 430) G x S = - 206400 S. The fact that $Gd_{36}Co_{64}$ film general behaviour is determined by Gd is confirmed by the sign of MOKE hysteresis loop.

Although the saturation magnetization of Co in multilayered structure can be less than that one of bulk Co (due to size effects and charge transfer effects) it can be taken as 1400 G for room temperature in first approximation. In this case for {Co(70 Å) /(Gd-Co)(120 Å)}$_4$/Co(70 Å) structure the total magnetic moment of Co layerss of square S can be written as 5 x 70 Å x 1400 G x S = 490000 S. Therefore total magnetization of {Co(70 Å) /(Gd-Co)(120 Å)}$_4$/Co(70 Å) structure is equal ($M_{Co}$ + $M_{GdCo}$)/830 Å = (490000- 206400)/830 Å = 340 G.

{Co(70 Å) /(Gd-Co)(120 Å)}$_4$/Co(70 Å) structure has two subsystems: five Co layers and four Gd-Co layers. When the temperature decreases the Co magnetization shows little change but magnetization of Gd-Co increases (Figure 4). We can estimate the compensation temperature as follow: (4×$M_{GdCo}$×120 Å)/830 Å = (5×$M_{Co}$×70 Å)/830 Å. If $M_{Co}$=1400 G, than $M_{GdCo}$ must be 1020 G in the compensation point. Figure 4 shows that this value of the magnetization corresponds to the temperature below 77 K.

Figure 4. Temperature dependence of magnetization of $Gd_{36}Co_{64}$ 1000Å thin film.

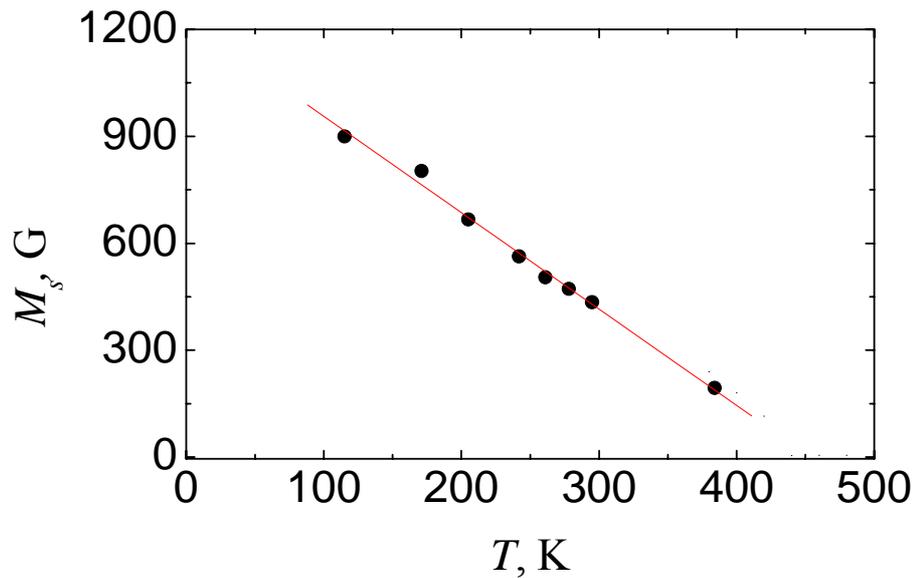

The confirmation of the fact that basic {Co(70 Å) /(Gd-Co)(120 Å)}$_4$/Co(70 Å) is a Co magntization dominated system was done by Kerr effect observation of the sign of hysteresis loops (mentioned in Table 1).

One may suppose that Co in pure Co layers is in general aligned with Co of GdCo layer because these two Co components are exchange coupled ferromagnetically and therefore Gd and Co for any Co containing layer are antialligned. Fig.3b shows schematically the VSM magetization curves and proposed magnetization process in such a structure. The behaviour up to the fields corresponding $H_{k1}$ was observed by VSM and corresponding data are mentioned in Table 1. At room temperature in a small fields below $H_{k1}$ all structure is collinear and pure Co layer moments are parallel to the external field, but the magnetic moments of Gd-Co layers are antiparallel to the external field. In a certain field denominated as $H_{k1}$ the Zeeman energy of Gd-Co layers becomes equal to the exchange interaction energy between layers of GdCo and Co. Non-collinear structure formation takes place with non-aligned position of two different Co moments (Co and Gd magnetization vectors are antialigned even in this first angular configuration). With an increase of the external field this angle become less and in second critical field $H_{k2}$ all magnetic moments become aligned. The dependence of the critical field $H_{k1}$ on the thickness of the non-magnetic fillers is clearly seen from data of Table 1.

We study the high frequency properties of {Co(70 Å) /(Gd-Co)(120 Å)}$_4$/Co(70 Å) structures with and without non-magnetic fillers of Si or Cu between magnetic layers.

**Powders**

The first type of powder was a commercial composite (Dynabeads M-450): 4.5 μm polystyrene beads consisting of a mixture of nanometer-sized iron oxides particles embedded in a polymer matrix (Fig. 5). The magnetic mass susceptibility of the Dynabead M-450 is 16±3 x 10$^{-5}$ m$^3$/kg. For FMR measurements they were dried (as the original material was in PBS suspension) and two types of new composite samples were prepared. The first one consisted of non-aligned more or less homogeneously distributed Dynabeads glued onto quartz substrates by GE7031 glue. In the second one the beads were allowed to set in the glue while being located in a permanent magnet (~ 1.7 kOe) thereby "aligning" the magnetic grains.

Figure 5. Optical microscope picture of M-450 magnetic Dynabeads of 4. 5 μm diameter.

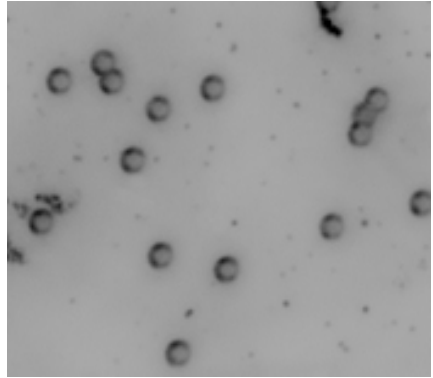

The second set of powders was nominally $Co_{80}Ni_{20}$ obtained by the polyol technique at Instituto de Ciencia de Materiales (Madrid, Spain) [10]. Table 2 gives some data on their magnetic properties and structure and Figure 3 shows the shape of the particles. Although the observed shape of the particles cannot be described very precisely because the images were taken at magnification not very far from the resolution limit of the KJ30 microscope there is no doubt that the shape of many particles tends to be hexagonal or at least one can discern hexagonal corners. It would be very risky to claim that these particles are spherical as appears to have been claimed in previous studies [13]. The data reported in literature [13] and preliminary X-ray study shows the presence of both hcp phase together with fcc phases in these materials.

Figure 6. $Co_{80}Ni_{20}$ 200 μm diameter nanoparticles without additional metal covering spread on a metallic plate. Scanning electron microscopy (observations done in collaboration with Dr. A. Ayari).

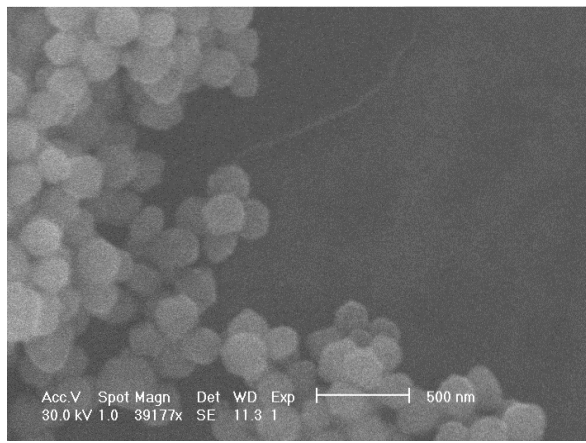

Table 2. Selected properties of $Co_{80}Ni_{20}$ powders used for FMR study. Center diameter $d_c$ means non-magnetic metal center used as a nucleating agent; d is a caliper dimension. Underlined numbers correspond to nanosize particles properties reported recently in [13]. The present particles are significantly softer magnets.

| d (nm) | $M_s$ (emu/g) | $M_r/M_s$ | | $H_c$ (Oe) | | $d_c$ (nm) |
|---|---|---|---|---|---|---|
| 200 | 150 | 0.015 | 0.09 | 40 | 170 | Ag of 4-5 nm |
| 100 | 130 | 0.08 | 0.17 | 200 | 280 | Pt of 2-3 nm |
| 45 | 120 | 0.28 | 0.44 | 440 | 570 | Pt of 2-3 nm |

For FMR measurements, the CoNi powders were first put in a vacuum grease. As we found some problems which may be caused by movement of the particles in high fields, subsequently non-aligned and aligned samples were prepared by the same method as described for Dynabeads. In order to check the influence of particular distributions of grains and their interaction levels, the preparation of aligned and non-aligned samples was repeated as well as an additional sample with a magnetic field applied perpendicular to the substrate was prepared. In this case particles formed agglomerates of a circular shape clearly seen by eye.

In addition, as the coercivity of the CoNi powders is rather high, in order to insure both the saturated state and alignment, samples were prepared in a high magnetic field of about 6 kOe.

3. **METHODS**

The detailed description of the microwave techniques based on usage of aconventional homodyne spectrometer has been reported in previous publications [5,12-16]. All FMR measurements in this study were done at room temperature.

To increase its range the electromagnet was completed by discs and calibrated for the whole range of magnetic fields from 0 to 15.5 kOe using Lake Shore 450 Gaussmeter and further checked by resonance of g-marker (g=2 for DPPH).

1. We used a microwave microscope, i.e. 2 mm diameter hole in a thinned wall of the cavity. The FeNi film was placed outside the cavity in front of the hole using a small amount of vacuum grease for fixing it at the ends. This gives us the advantages of i) avoiding an extra large load, ii) estimating the homogeneity of the properties by exposing different regions of the sample to microwaves, and iii) making measurements from two sides of the sample: from the side of the film and from the side of the substrate (as the glass substrate is not opaque to microwaves).
2. Several cavities operating between 9.5 and 11.8 GHz were used.
3. When the samples were set inside the cavity it was possible to measure angular dependence of FMR for the external field rotated not only in plane perpendicular to the film, [sample on wall (Figure 7)] but also in plane of the film, [sample at bottom].

The angular dependence of the resonance field, $H_{res}$, was used to establish that the magnetization was uniform and that the strain induced anisotropy field had a symmetry axis along the film normal [16]. Namely, $H_{res}$ is given by the equation:

$$\left(\frac{\omega}{\gamma}\right)^2 = (H_{res}\cos\alpha - 4\pi M_{eff}\cos\theta)^2 + H_{res}\sin\alpha\,(H_{res}\sin\alpha + 4\pi M_{eff}\sin\theta) \quad (1\text{ a})$$

combined with the equilibrium condition

$$\left(\frac{\sin(\theta-\alpha)}{\sin\theta\cos\theta}\right) = \frac{4\pi M_{eff}}{H_{res}} \quad (1\text{ b})$$

where $\omega = 2\pi f$ and $\theta(\alpha)$ measures the angle between **M(H)** and the film normal.

Figure 7. Geometry of the angular dependence of FMR measurements in thin films. Spin wave resonance modes can be observed in thin films at certain conditions for $\alpha=0^\circ$

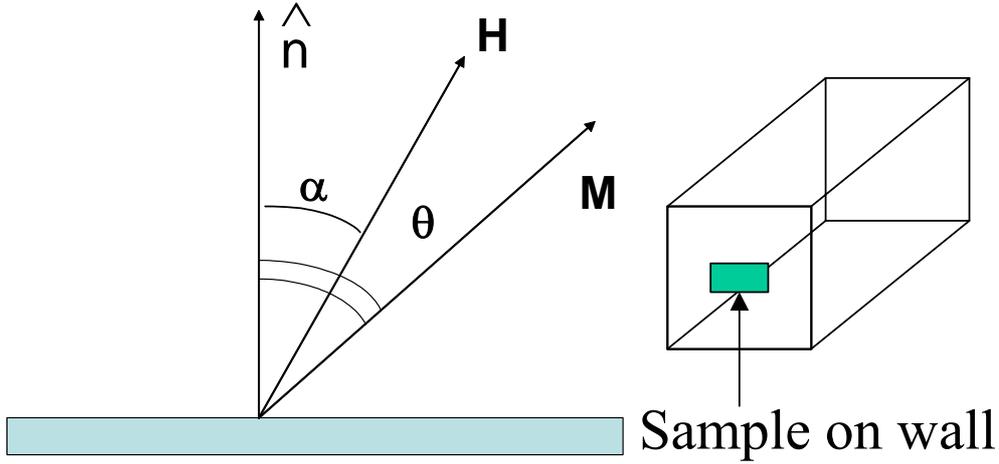

4. For some thin films in the **H** $\perp$ film configuration, a set of spin wave resonances (SWR) were observed. In the simplest case the resonance fields should vary as the square of the mode number, n, and the slope $\left(\frac{\partial H_n}{\partial n^2}\right) = 2\pi^2 A/(ML^2)$, were A is the exchange stiffness, M the static magnetization and L is a film thickness. As shown in the Figure 8 we were able to identify SWR modes up to n = 9 for FeNi film and obtain for A the value of $1.2 \times 10^{-6}$ erg/cm in excellent agreement with previously published data.

5. In case of powder samples FMR measurements and measurements of the microwave losses were done in the geometry described in figure 6 and the field rotated from H $\perp$ $h_{rf}$ for FMR to H $||$ $h_{rf}$ for magnetoabsorption. For the sample located at the bottom of the cavity $e_{rf} = 0$; $h_{rf}$ = max.

Figure 8. Spin wave resonance fields for Permalloy thin film.

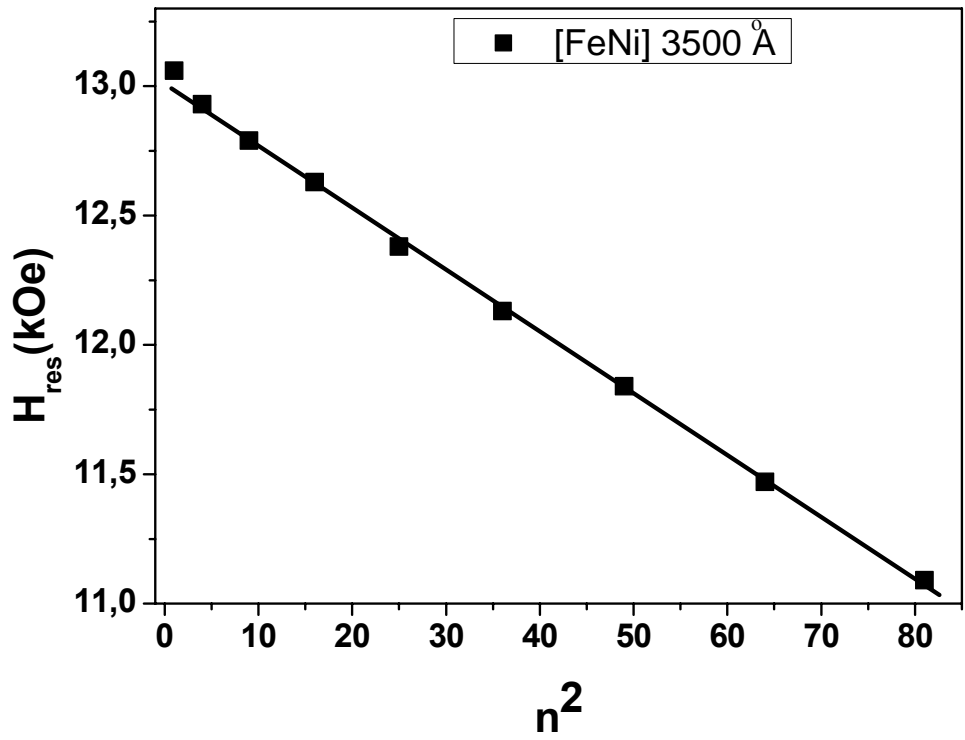

Figure 9. Geometry of the FMR and microwave losses measurements for powder samples. When sample is on the bottom of the cavity it is in $h_{rf}$ field, but when it is in the middle of the cavity it is in a $e_{rf}$ field.

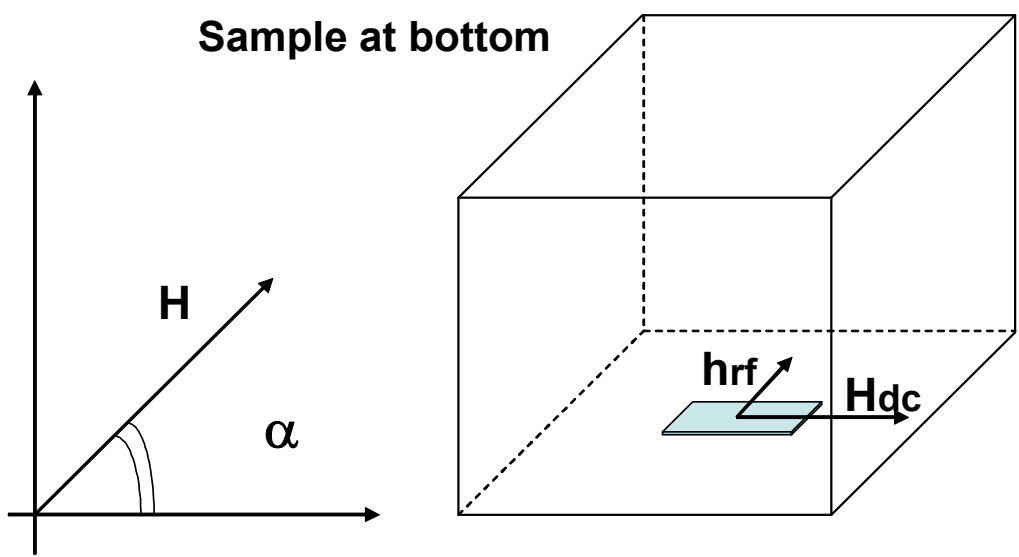

6. Microwave magnetoabsorption for CoNi powders was also measured when the sample is shifted to the region of the microwave electrical field $e_{rf}$, i.e. in the middle of the $\lambda/2$ cavity. As before data on magnetoabsorption can be represented by the following function [12]:

$$f(H) = \left( \frac{P(0) - P(H)}{P(H)} \right) \qquad (2)$$

where P(H) is the field-dependent microwave absorption.

## 4. RESULTS

**MI Permalloy series**

### A. Single layered fillms

Single layered FeNi films were measured with microwave microscope as well as inside the cavity using different pieces of the same film. In the parallel geometry ($\alpha = 90°$) essentially one resonance line was observed being about 80 Oe wide for the microscope case and about 140 Oe for measuremenst inside the cavity. The resonance fields lie between 1.14 -1.18 kOe indicating that the films are reasonably uniform (see Table 3). At the same time since the in-plane orientation is arbitrary it demonstrates that the anisotropy axis, if any, is normal to the film. The 500 Å film (obtained from Rowan University) however, yielded a significantly different value for $H_{res}$ at $\alpha = 0$.

For uniaxial anisotropy where the symmetry axis is along the film normal one can write $4\pi M_{eff} = 4\pi M + \left( \frac{2K_u}{M} \right)$ and obtain from equations (1a), (1b) that

$$\left( \frac{\omega}{\gamma} \right)^2 = H(H + 4\pi M_{eff}) \qquad \mathbf{H} \parallel \text{film} \qquad (\alpha = 90°) \qquad (3a)$$

$$\left(\frac{\omega}{\gamma}\right) = H - 4\pi M_{eff} \qquad \mathbf{H} \perp \text{film} \qquad (\alpha = 0^o) \qquad (3b)$$

Equations (3a,b) can be used to determine $4\pi M_{eff}$ and thereby obtain the angular dependence implied by Eqs (1a,b) shown in figure 10 for 1-753 film, where we have also plotted the experimental results. The close agreement is quite satisfactory. However, $4\pi M_{eff}$ is significantly different from $4\pi M$ (10.05 kOe) implied by SQUID data (Fig.2). The results are collected together in Table 3.

Figure 10. Theoretical and experimental angular dependencies of the resonance field of $Ni_{80}Fe_{20}$ thin film.

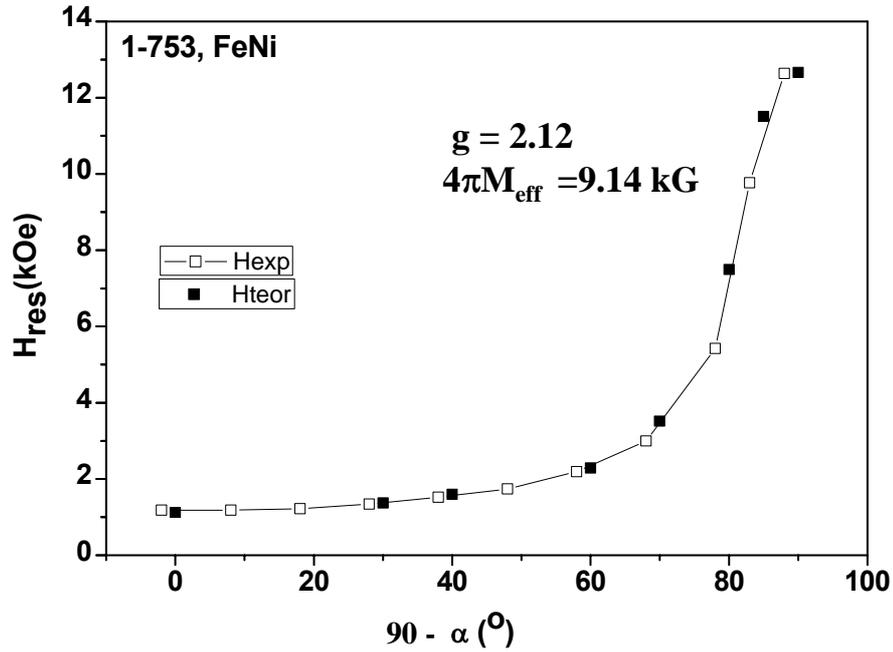

**B. Multilayers**

In this case spectra often consisted of more than one line. For 1-752, which was prepared without opening the chamber ( see Table 3) two lines are seen with the microscope when the film faces the opening but only one line appears when the glass is toward the opening. It is notable that this line has the same $H_{res}$ for $\alpha = 90^o$ as for the single layered film. So it seems reasonable to claim that it comes from the FeNi layer closest to the glass substrate. The second line is at a significantly higher field and yields a highly reduced value of $4\pi M_{eff}$. Most likely, this is because Cu diffuses into the FeNi layer thereby suppressing $M_{eff}$.

However when 1752 was placed inside and at the bottom of the cavity it yielded a spectrum consisting of 4 only partially resolved lines with a weak dependence on orientation of the applied field. The out-of plane angular variation (Fig. 8) is again in accord with equations 1(a) and 1(b). So that in this case also the anisotropy axis must be along the normal. The SWR spectrum exhibits only 4 resonance lines but the A value ($1.25 \times 10^{-6}$ erg/cm )is quite reasonable.

If the chamber is opened between depositions of the layers of the multilayered structure we obtain the film designated 1-754. In this case one observes essentially the same result (Fig. 12) as for single layered film except that even fewer SWR modes are discernible and the A value, although less well determined, appears to be suppressed. Again, it seems reasonable to to claim that opening the chamber causes formation of the oxide which may inhibit the diffusion of Cu involved to describe the results for 1752.

As seen from Table 3 the more complex GMI-multilayer geometry (sample 1-755) has little or no influence on the FMR spectra compared with data for 1-754 sample.

Figure 11. Theoretical and experimental angular dependencies of the resonance field of $Ni_{80}Fe_{20}$ /Cu/ $Ni_{80}Fe_{20}$ multilayer deposited without opening the vacuum chamber. Measurements done with microscope both from the glass and the film sides (accordingly one and two resonances respectively were observed), f = 10.45 GHz, room temperature.

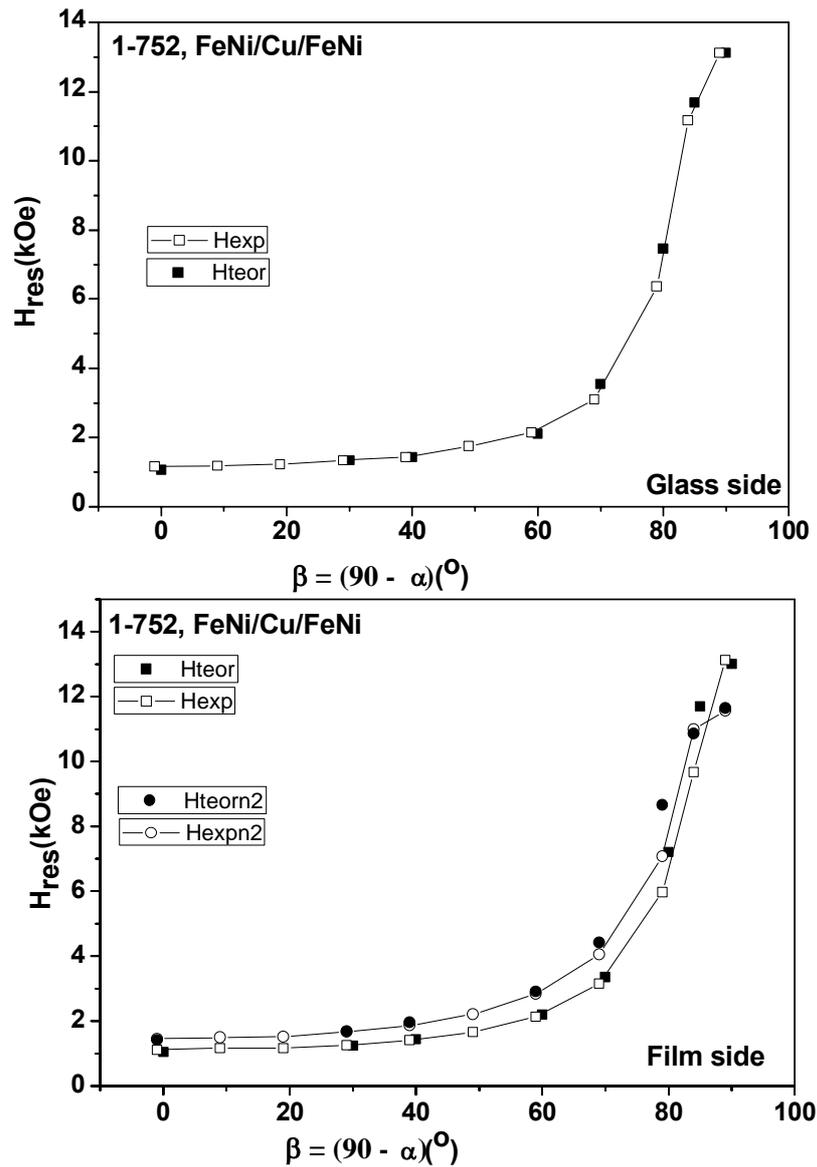

Figure 12. Theoretical and experimental angular dependencies of the resonance field of $Ni_{80}Fe_{20}$ /Cu/ $Ni_{80}Fe_{20}$ part of multilayer deposited with opening the vacuum chamber, f = 10.45 GHz, room temperature.

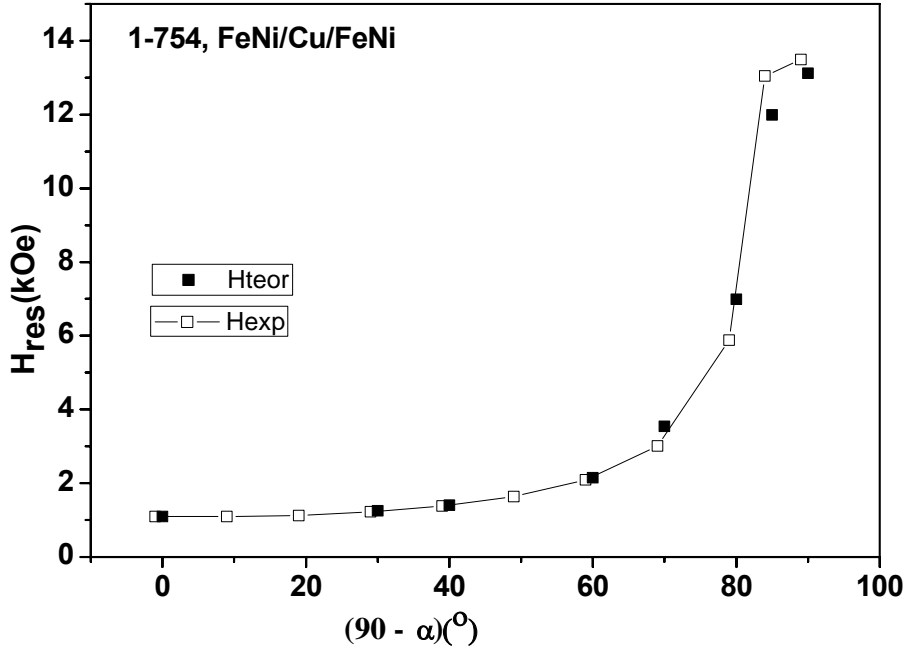

Table 3. Selected data on FMR at f = 10.45 GHz in $Ni_{80}Fe_{20}$ thin films and Permalloy based structures. Resonance fields are done for $\alpha = 0$ and $90°$.

| Sample | $H_{res1}$ kOe $\alpha=90$ | $H_{res1}$ kOe $\alpha=0$ | $H_{res2}$ kOe $\alpha=90$ | $H_{res2}$ kOe $\alpha=0$ | $4\pi M_1$ kG | $g_1$ | $4\pi M_2$ kG | $g_2$ | $\left(\dfrac{\partial H_n}{\partial n^2}\right)$ (n, number of SWR peaks) | A erg/cm $10^{-6}$ |
|---|---|---|---|---|---|---|---|---|---|---|
| 1-753 FeNi | | | | | | | | | | |
| A: microscope | 1.18 | 12.64 | - | - | 9.14 | 2.12 | - | - | - | - |
| A : inside the cavity | 1.14 | 13.06 | - | | 9.68 | 2.18 | - | - | 5 | |
| *B: inside the cavity | 1.14 | 13.20 | - | | 9.56 | 2.04 | - | - | 24 (n = 9) | 1.2 |
| 1-752 FeNi/Cu/FeNi | | | | | | | | | | |
| microscope Film side | 1.11 | 13.00 | 1.46 | 11.56 | 9.56 | 2.16 | 7.87 | 2.01 | 99 (n = 4) | 1.25 |
| Glass side | 1.16 | 13.11 | - | - | 9.58 | 2.10 | | | | |
| 1-754 FeNi/FeNi | 1.06 | 13.12 | 1.12 | 13.12 | 9.74 | 2.19 | 9.65 | 2.14 | - | - |
| FeNi/Cu/FeNi | 1.10 | 13.49 | - | - | 10.0 | 2.13 | - | - | 61.3 (n = 3) | 0.7 |
| 1-754 FeNi{Cu}FeNi FeNi/FeNi microscope | 1.06 | 13.33 | 1.13 | - | 9.92 | 2.18 | No 90° | | 61(n = 4) | 0.7 |
| microscope | 1.12 | 13.25 | 1.18 | - | 9.76 | 2.13 | | | 50 | 0.6 |
| FeNi/Cu/FeNi microscope film to cavity | 1.13 | 13.42 | - | | 9.89 | 2.10 | - | - | 50 (n = 4) | 0.6 |
| microscope glass to cavity | 1.13 | 13.01 | - | | 9.54 | 2.14 | - | - | - | - |
| 1260 onto Si mono SiO$_2$/TiN/FeNi/TiN/FeNi/ TiN/Cu/TiN/FeNi/ TiN/ FeNi film to cavity | 1.23 | 12.23 | - | - | 8.73 | 2.12 | - | - | - | - |
| FeNi (Prof. S. Lofland sample) | 1.4 | 11.26 | - | - | 7.69 | 2.09 | - | - | - | - |

The slightly higher resonance field and lower magnetization in case of the 1260 complex multilayer designed with the idea of clear electrical separation of FeNi components from the conducting central part makes us think that internal stresses were increased in the system, at least one may see clear disadvantage of it from the magnetic state point of view. The line width of 120 Oe and especially the complex shape with perhaps three non-resolved lines indicate that in such a manufactured structure magnetic inhomogeneities are rather significant.

**C: Co/GdCo multilayers**

**FMR in single layered films**

Co/GdCo multilayers (see Table 1) are complicated systems consisting of different magnetic and non-magnetic components. In order to develop reasonable interpretations of the FMR data in multilayers it was necessary to first study the resonance behaviour of single layered magnetic films of pure Co and $Gd_{36}Co_{64}$ of the same composition as the components of the multilayered structure and prepared in similar conditions.

1199 Co film resonance field Angular dependence of FMR in 1199 Co film was first measured with the sample at the bottom of the cavity in order to study the in-plane magnetic anisotropy. The resonance field varied by less than 0.04 kOe indicating absence of in-plane magnetic anisotropy.

Angular dependence in the perpendicular plane for the same film was measured in the interval of $5 \leq \alpha \leq 90^o$ (Fig. 4). The resonance consists of one line of about 100 Oe width for $\alpha = 90^o$ orientation. Since it was not possible to measure $H_{res}$ for $\alpha = 0^o$ we assumed g = 2.2 ($\omega/\gamma \cong$ 3.39) in order to obtain $4\pi M_{Co} \cong 14.8$ kG ($M_{Co} \cong 1180$ G) which is much smaller than the buk value of 17.6 kOe. Most likely explanation is a significant contribution from ($2K_u/M$) although some deficit could also be due to small thickness

Two pieces 1-703-1and 1-703-2 of GdCo single layer film were measured when sample was inside the cavity and on the cavity wall. Two resonance lines were observed at $\alpha = 90^o$ in both cases. For $\alpha = 0^o$ two lines were observed in 1-703-2 but a wide "line" consisting of more than one unresolved resonance appeared in 1-703-1. Reasonably close values: $4\pi M_{GdCo}$= 3.5 kOe (g=1.80; $M_{eff}$ = 275 G ) for 1-703-1 and 3.8 kOe (g=1.82; $M_{eff}$ = 300 G) for 1-703-2 were obtained. Figure 13 shows the angular dependence for $\alpha = 5$ to $90^o$ for Co and for $\alpha = 90$ to $0^o$ for two $Gd_{36}Co_{64}$ films for comparison.

Figure 13. Angular dependence of the resonance fields observed in Co and $Gd_{36}Co_{64}$ thin films at f = 10. 45 GHz, room temperature.

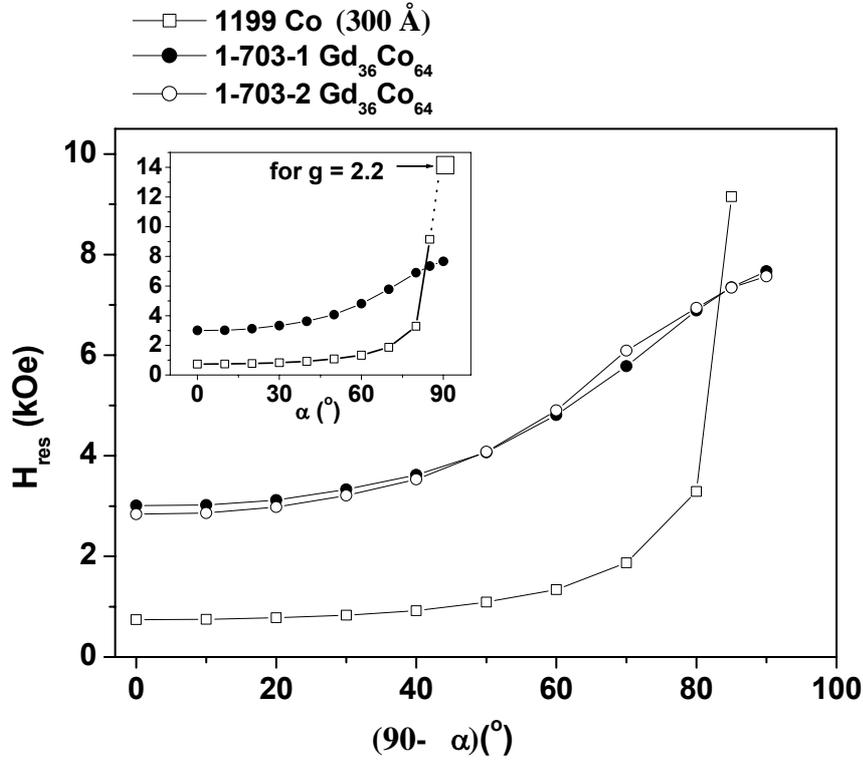

The value of $M_{eff}$ calculated from our FMR data is less than the effective magnetization reported for GdCo films films by other authors ($4\pi M$ = 5 kOe) [17] of the same composition. One of the possible explanations can be that our GdCo thin film was covered by Si layer of more than 1000 Å thickness in order to avoid oxidation. It was shown recently [2] that Si can be a problematic material because it can influence Co by charge transfer and can decrease the cobalt magnetization. The decrease of Co magnetization can cause the Gd magnetization to decay because the Gd sublattice has its own Curie temperature close to the room temperature and as a consequence cause reduction of the magnetization of GdCo film as a whole.

At the same time the value of $M_{eff}$ calculated from our FMR data is less than the M measured by rotational anisometer for room temperature (Figure 14). This discrepancy can be explained if we suppose that the relatively thick GdCo film (Gd$_{0.36}$Co$_{0.64}$ film of 1000 Å) has a perpendicular anisotropy component which should be taken into account for the analysis of FMR data.

Figure 14. Temperature dependence of DC magnetization for Gd$_{0.36}$Co$_{0.64}$ thin film of 1000 Å thickness.

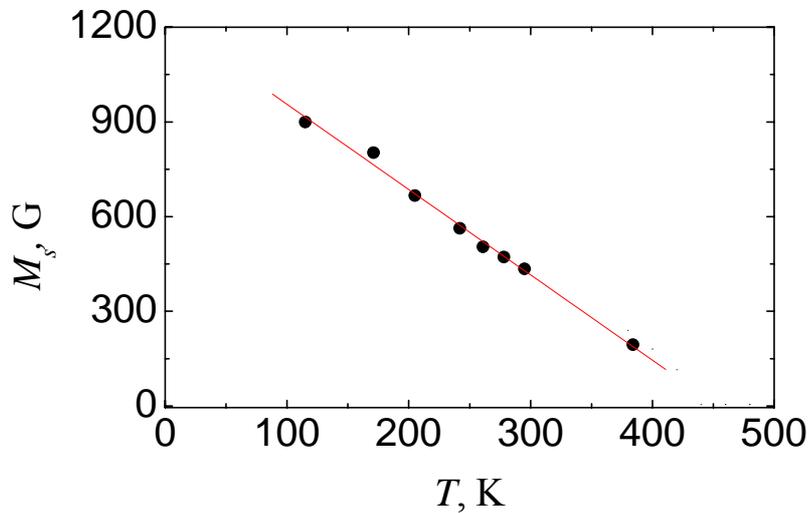

### **FMR in Co/GdCo multilayers**

As shown in Figure 15 angular dependence (sample on the cavity wall) of resonance field was measured in 1-694 multilayered structure of {Co(70 Å) /(Gd-Co)(120 Å)}$_4$/Co(70 Å) for the interval of $5 \leq \alpha \leq 90°$. The resonance consists of one line which is about 280 Oe wide for $\alpha = 90°$ orientation. Since the resonance field in the perpendicular orientation was not accessible we put g = 1.8 (see page 23 for g values for GdCo film) which is reasonable for GdCo system of composition under consideration ($\omega/\gamma \cong 4.14$) in order to calculate $4\pi M \cong 13.1$ kOe (M$_{eff} \cong 1050$

G). If we use g = 2.2 which is more reasonable for Co ($\omega/\gamma \cong 3.38$) $4\pi M \cong 8.60$ kOe ($M_{eff} \cong 680$ G).

Another possible interpretation may be that since the applied field is low, $M_{Gd-Co}$ is antiparallel to $M_{Co}$ and one can treat the composite film as a two sublattice *ferrimagnet* in which case the resonance field in parallel geometry ($\alpha = 90°$) is given by (for acoustic mode)

$$\left(\frac{\omega}{\gamma_{eff}}\right)^2 = H\,[H + 4\pi(M_{Co} - M_{Gd-Co})] \tag{4}$$

where $\gamma_{eff} = \dfrac{M_{Co} - M_{Gd-Co}}{\dfrac{M_{Co}}{\gamma_{Co}} - \dfrac{M_{Gd-Co}}{\gamma_{Gd-Co}}}$ ; $M_{Co} = 1180$ G, $M_{Gd-Co} = 300$ G (from FMR data p.23).

Using g = 2.2 for Co and 1.8 for Gd-Co ; $g_{eff} = \dfrac{1180 - 300}{\dfrac{1180}{2.2} - \dfrac{300}{1.8}} = 2.38$

so $\dfrac{\omega}{\gamma_{eff}} = 3.14$ and yields $(4\pi M_{Co} - 4\pi M_{Gd-Co}) = 7.17$ kOe ($M_{eff} \approx 570$ G for $H_{res} = 1.18$ kOe) which is also much larger than the measured value. Unfortunately, the discrepancy is equally severe.

Figure 15 shows for comparison the angular dependences for $90 \leq \alpha \leq 5°$ (sample on the cavity wall) corresponding to Co, GdCo and {Co(70 Å) /(Gd-Co)(120 Å)}$_4$/Co(70 Å) system. In complicated systems like multilayers it is important to make sure that the observed properties are repeatable, i.e. really correspond to the structure under consideration and are not caused by uncontrolled additional parameters. Before studying the influence of the thickness of different non-magnetic layers on the high frequency properties of the {Co(70 Å)/Cu(10 Å)/(Gd-Co)(120 Å) )/Cu(10 Å)}$_4$/Co(70 Å) structure, we first checked for in plane anisotropy by using angular dependence measurements when the sample was at the bottom of the cavity. The observed variation in resonance field was less than 2% indicating absence of in-plane magnetic anisotropy.

Figure 15. Angular dependence of resonance field for selected Co-containing samples.

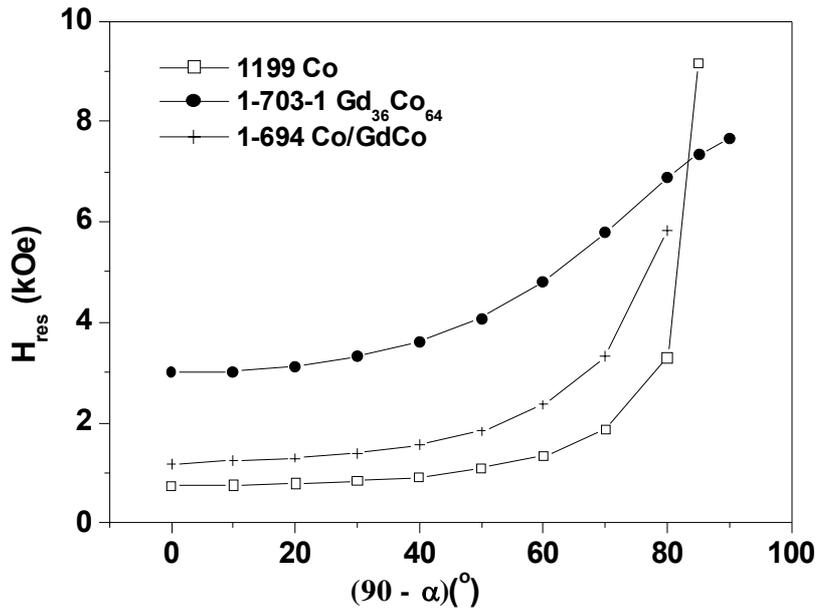

The FMR fields and line widths were measured for several different pieces (each of order of 15 mm$^2$ square) of 1-721 (Table 4) prepared by the same sequence of deposition protocol. The variations notable in Table 4 indicate that different pieces of the same film are far from identical and this fact should be taken into account in any analysis of the data for the samples with non-magnetic layers.

Table 4. Resonance fields of {Co(70 Å)/Cu(10 Å)/(Gd-Co)(120 Å) )/Cu(10 Å)}$_4$/Co(70 Å) sample for $\alpha = 90^o$ orientation, f = 10.4 GHz using g = 2.2

| Sample | Number of resonance lines | $H_{res1}$, kOe | $H_{res2}$, kOe | Width, Oe |
|---|---|---|---|---|
| 1-721-2 | 1 | - | 0.84 ($4\pi M_s$= 12.7 kOe) | 115 |
| 1-721-3 | 1 | - | 0.91 ($4\pi M_s$= 11.6 kOe) | 220 |
| 1-721-4 | 2 | 0.68 ($4\pi M_s$= 16.1 kOe) | 0.88 ($4\pi M_s$= 12.1 kOe) | - |
| 1-721-5 | 1 | - | 0.77 ($4\pi M_s$= 14.1 kOe) | 130 |
| 1-721-6 | 1 | - | 0.83 ($4\pi M_s$= 12.9 kOe) | 160 |

**{Co(70 Å)/Si/(Gd-Co)(120 Å) )/Si}$_4$/Co(70 Å) multilayers**

Figure 16 collects data on FMR in Co/GdCo based multtilayers with Si non-magnetic sublayers of different thicknesses. For all samples FMR was measured in a wide field interval from zero to fields higher than that corresponding the resonance field for Gd$_{36}$Co$_{64}$. The multilayers with Si thicknesses of 10 and 7 Å show two well resolved resonances. The higher field resonance in each {Co(70 Å)/Si/(Gd-Co)(120 Å))/Si}$_4$/Co(70 Å) sample appears at a field very close to that observed in pure GdCo film (Figure 16 a), i.e. when Si thickness is ≥ 7 Å the GdCo layer inside the multilayered film seems to be unaffected by the presence of other components.

In all samples containing Si sublayers the low field resonance was relatively close to the Co film resonance. This means that in each {Co(70 Å)/Si/(Gd-Co)(120 Å) )/Si}$_4$/Co(70 Å) the Co layer is relatively undisturbed by other sublayers although further study may reveal better quantitative information. As seen from Figure 17 the present data show a rather weak effect. One additional point has to be mentioned. For 1-709 sample with a 2 Å Si sublayer very "lossy" behaviour was observed. The signal was so weak that we were able to measure only $\alpha = 90°$. It seems that the most inhomogeneous state corresponds to the 2 Å Si sublayer.

Figure 16. Angular dependence of FMR field for {Co(70 Å)/Si/(Gd-Co)(120 Å) )/Si}$_4$/Co(70 Å) multilayers and complementary Co and GdCo films, f = 10.45 GHz.

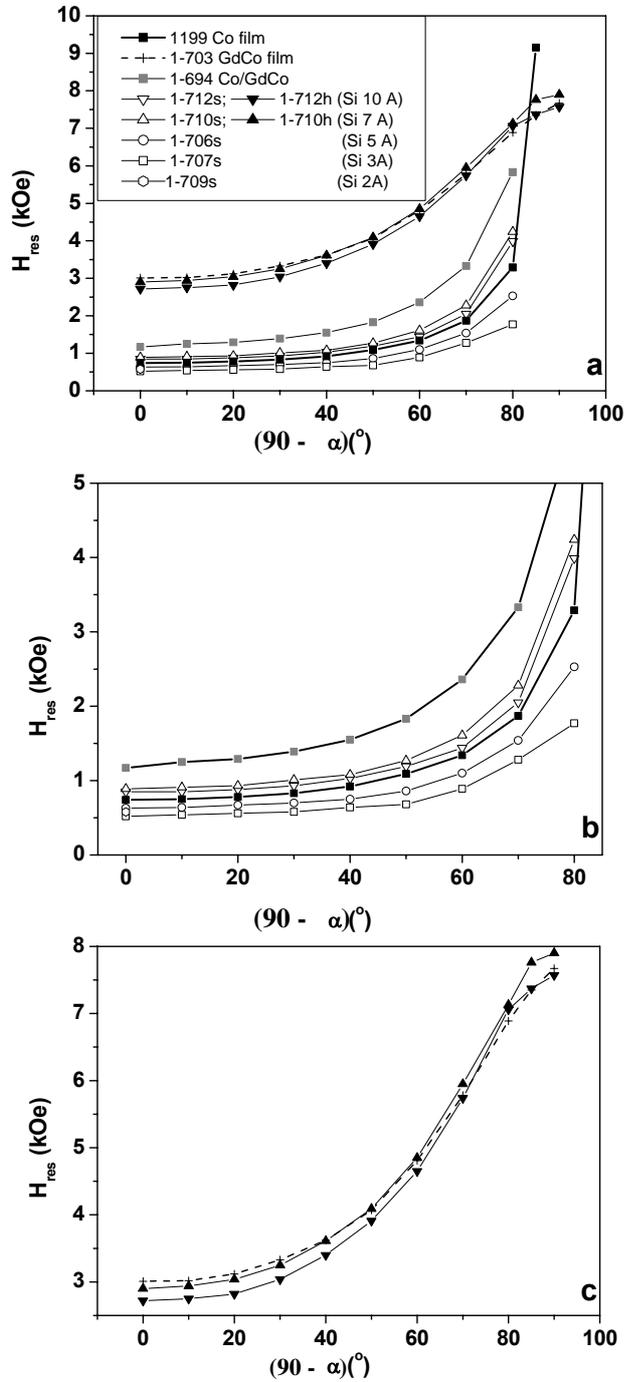

Figure 17. Dependence of the resonance field of {Co(70 Å)/Si/(Gd-Co)(120 Å) )/Si}$_4$/Co(70 Å) multilayers on the thickness of Si sublayer for $\alpha = 90^o$, f = 10.45 GHz.

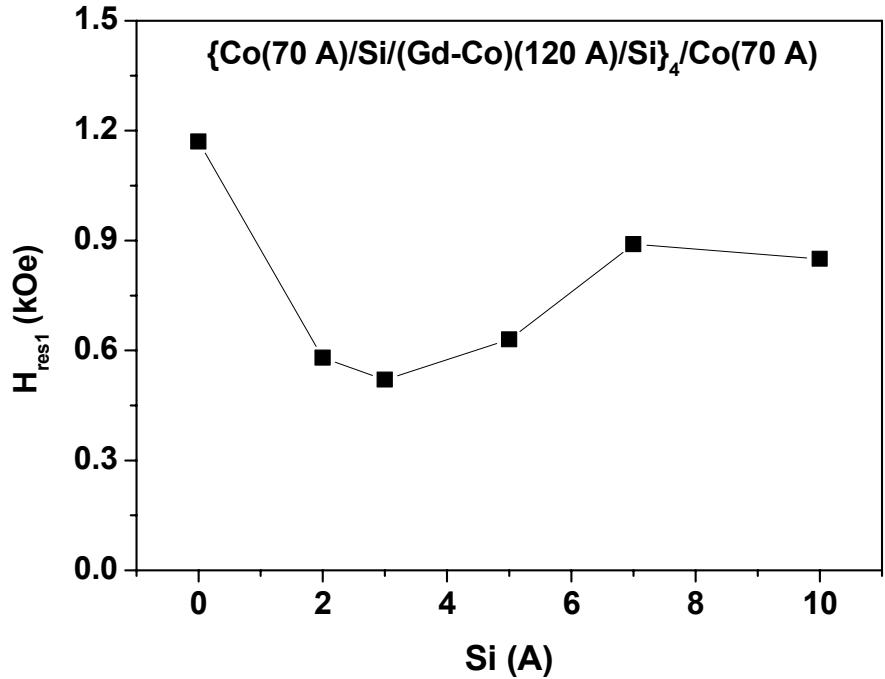

**{Co(70 Å)/Cu/(Gd-Co)(120 Å) )/Cu}$_4$/Co(70 Å) multilayers**

Figure 18 and Table 5 collect data on FMR in Co/GdCo based multilayers with Cu non-magnetic sublayers of different thicknesses. The measurements covered a wide field interval from zero to fields higher than the resonance field for Gd$_{36}$Co$_{64}$ sample but only in one case [1-723 multilayer with thickness of Cu of 15 Å] two resonances in different fields ( one close to Co and another close to GdCo) were observed.

High field resonance appeared for {Co(70 Å)/Cu/(Gd-Co)(120 Å) )/Cu}$_4$/Co(70 Å) sample in the field very close to that of pure GdCo film (for $\alpha = 0^o$ H$_{res}$ = 2.96 kOe) i.e. at a high thickness of Cu GdCo layer inside the multilayer seems non-affected by presence of other components of multilayer.

Figure 18. Angular dependence of resonance field for {Co(70 Å)/Cu/(Gd-Co)(120 Å) /Cu}$_4$/Co(70 Å) multilayers and complementary Co film, f = 10.45 GHz, room temperature.

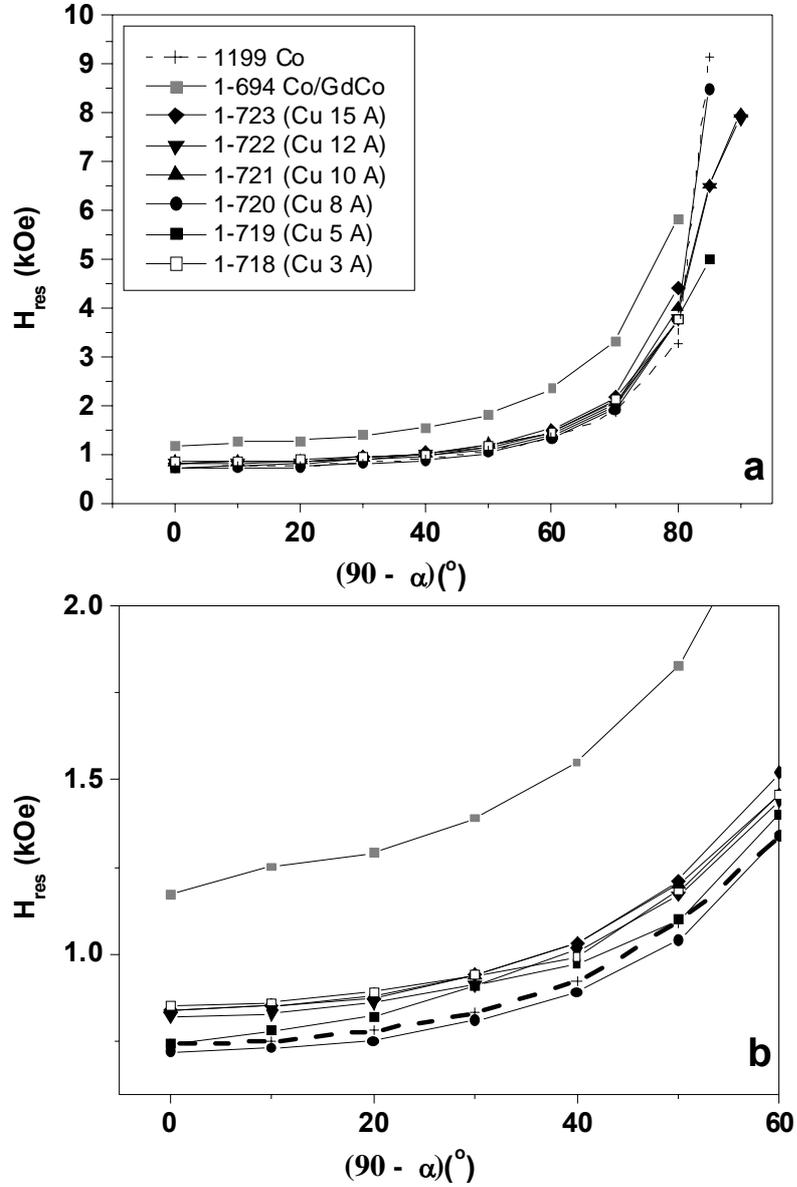

Table 5. Resonance fields of {Co(70 Å)/Cu/(Gd-Co)(120 Å) )/Cu(10 Å)}$_4$/Co(70 Å) sample for $\alpha = 0°$ orientation, f = 10.4 GHz in supposition g = 2.2

| Sample | Number of lines $\alpha = 90^\circ$ | $\alpha = 10^\circ$ | Cu thickness Å | $H_{res1}$, kOe | $4\pi M_s$ kOe | Width, Oe |
|---|---|---|---|---|---|---|
| 1-694 | 1 | 1 | 0 | 1.17 | 8.58 kOe | 317 |
| 1-718 | 2 | 2 | 3 | 0.85 | 12.57 kOe | doublet |
| 1-719 | 1 | 2 | 5 | 0.74 | 14.68 kOe | 460 |
| 1-720 | 1 | 2 | 8 | 0.72 | 15.13 kOe | 95 |
| 1-721 | 1 | 1 | 10 | 0.84 | 12.75 kOe | 115 |
| 1-722 | 1 | 1 | 12 | 0.82 | 13.09 kOe | 135 |
| 1-723 | 1 | 2 | 15 | 0.84 | 12.75 kOe | 100 |

Figures 19-20 summarize the data as a function of Cu thickness.

Figure 19. Influence of the thickness of non-magnetic sublayer on FMR field for {Co(70 Å)/Si or Cu/(Gd-Co)(120 Å) )/Si or Cu}$_4$/Co(70 Å) multilayers, f = 10.4 GHz, $\alpha = 90^\circ$ room temperature.

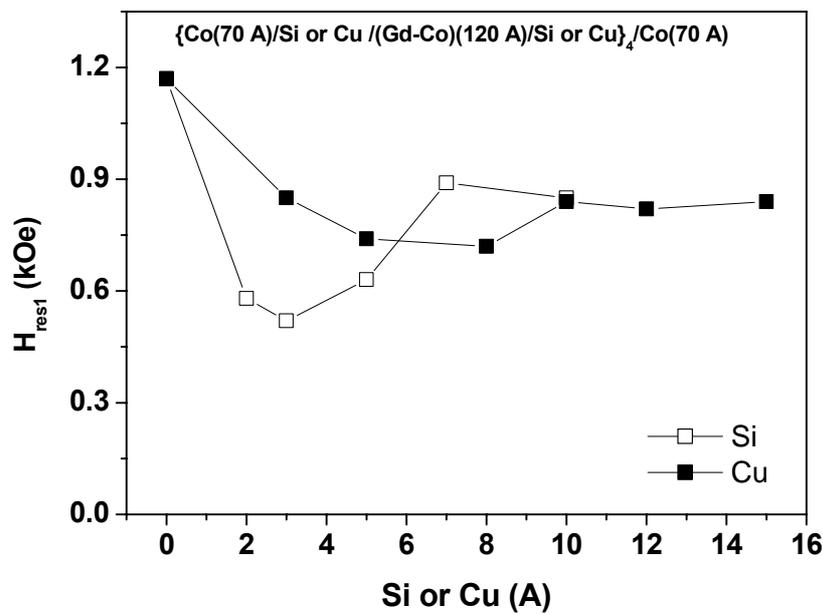

Figure 20. Influence of the thickness of non-magnetic sublayer on the width of FMR line of {Co(70 Å)/Cu/(Gd-Co)(120 Å) )/Cu}$_4$/Co(70 Å) multilayers

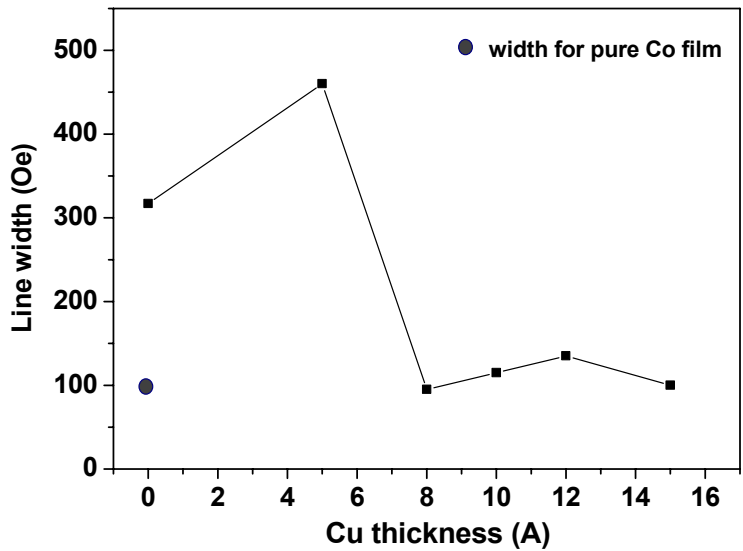

In Si set of samples 1-706 and in Cu set of the samples 1-720 multilayers show peculiar behaviour in a very low field as reflected by the unusual shape of the resonance curves. These additional lines are wide and clearly there are "zero" field signals in the data (Figures 21-22).

Figure 21. FMR signals for sample 1-706 ("zero" field signal is present).

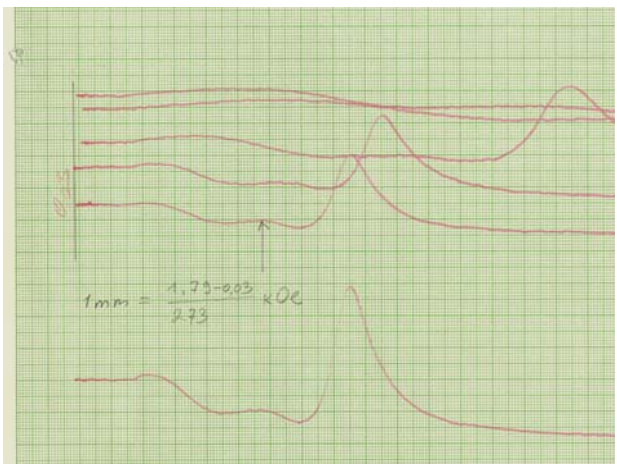

Figure 22. FMR signals for sample 1-720 ("zero" field signal is present).

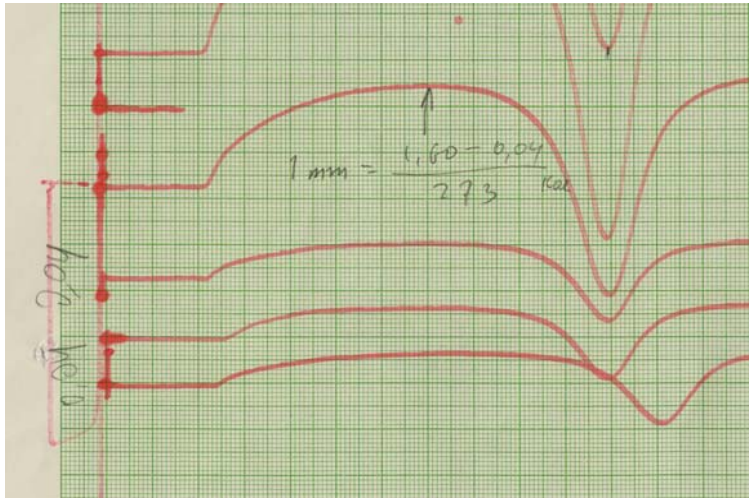

### D: Powder based systems

#### a) Dynabeads

The FMR was measured for a non-aligned sample (Fig. 9 geometry) for 3 different amounts of material held in the glue. Varying the packing of the beads had little effect on values of $H_{res}$ of 3.2 and 3.4 kOe. The linewidth is 400 Oe with a line shape close to Lorentzian. For the aligned sample the position of the peak was about 3.1 kOe with roughly the same width. Taking $H_{res} \cong 3.2$ kOe for all cases implies that the g-value is about 2.3. Since this is somewhat high for an iron oxide (g = 2.1) it is likely that there is an effective anisotropy field of about 300 Oe augmenting the external field.

Such a broad peak is not surprising for powders because deviations from sphericity can cause a wide distribution of demagnetising fields and also in a multicrystal the distribution of anisotropy fields can be sizeable.

#### b) CoNi powders

The second powder system was CoNi (with 3 polycrystalline samples of different particle sizes all of them large enough to accommodate more than one domain, see Table 2). The measured resonance fields of non-aligned samples are shown in Table 6. In all cases huge absorption in small fields indicating presence of a microwave magnetoimpedance effect made measurements rather difficult because of the superposition of two signals in the same field range. Therefore both FMR and field dependence of microwave losses were measured. Due to influence of the magnetoimpedance effect and rather broad lines it was difficult to obtain precise values for the line width.

The FMR measurements show that resonance fields (see Table 6) are unlikely to be due to spherical particles. Indeed in order to account for the observations one has to assume that the grains are diskettes with the H field parallel to the plane. If so, the following equation can be written $\left(\dfrac{\omega}{\gamma}\right)^2 = (H + H_k)(H + H_k + 4\pi M_s)$. The $H_k$ values were chosen to yield the same value of g in every case (Table 6). The increase in $K_1$ with reducing d is in accord with the observation that the coercive field increases as d is reduced.

However, further investigation on these powders indicate that one can not treat this analysis too seriously. As seen from Table 7 not only the aligned powders but also the non-aligned samples can exhibit sizeable angular variations. This is most likely due to formation of agglomerates or clusters which produce significant local dipolar fields. Powders are invariably problematic for FMR studies, but non-spherical grains combined with a large magnetocrystalline anisotropy effect literally frustrate any attempt of using this technique to study the available nanoparticles.

The zero field absorption and concomitant magnetoabsorption discussed next are interesting in their own right.

Table 6. Some data on FMR at f = 10.42 GHz in CoNi powders with different average size of the particles, where $H_a = \left(\dfrac{2K_1}{M}\right)$.

| d (nm) | $M_s$ (emu/g) from SQUID | $4\pi M$ (kOe) | $H_{res}$ (kOe) | $H_a$ (kOe) | g | $K_1$ (erg/cm$^3$) |
|---|---|---|---|---|---|---|
| 200 | 150 | 16.0 | 1.54 | 0.75 | 2.04 | $4.84 \cdot 10^5$ |
| 100 | 130 | 13.9 | 1.80 | 0.90 | 2.04 | $4.97 \cdot 10^5$ |
| 45 | 120 | 12.8 | 2.04 | 1.1 | 2.05 | $5.61 \cdot 10^5$ |

Table 7. Some data on FMR in CoNi powders with different average size of the particles. *

| d (nm) | State | $H_{res}$ for 0° (kOe) | $H_{res}$ for 90° (kOe) | f (GHz) |
|---|---|---|---|---|
| 200 | | | | |
| | Non-aligned A | 2.27 | 1.34 | 11.875 |
| | Non-aligned B | 2.32 | 1.32 | 11.875 |
| | Aligned perp. to plane* | 1.81 | 2.11 | 9.5 |
| | Aligned in a field | 2.38 | 1.14 | 10.42 |
| | Aligned in a field | 2.20 | 1.10 | 9.5 |
| 100 | Aligned in a field | 1.72 | 0.74 | 9.5 |
| | Aligned in a high field | 1.55 | 0.95 | 9.5 |
| | Non-aligned | 2.15 | 1.20 | 11.875 |
| 45 | Non-aligned | 2.12 | Below 0 | 9.5 |
| | Non-aligned | 2.30 | Below 0 | 11.875 |
| | Aligned in a field | 2.06 | Below 0 | 9.5 |

We have studied the magnetic field dependence of the microwave losses in all samples. In order to check that this effect is a magnetic one 200nm powder was also measured in $e_{rf}$ field. Although a very small change of f(H) was observed this variation is most probably due to the difficulty to put a large sample precisely in a homogeneous zone only electrical field (figure 23). For particles of all sizes there is a large zero field microwave absorption, which disappears on application of a field parallel to the rf magnetic field. For certain field interval f(H) shows linear dependence with a slope of about 0.167 $Oe^{-1}$ followed by very fast saturation.

We have roughly estimated the skin depth as $\delta \approx$ 1-2 μm (assuming the resistivity of CoNi of this composition is around 10 μΩcm). This means that the skin depth for the particles of all the sizes under consideration is significantly larger then their calliper size .

Figure 23. External magnetic field dependence of f(H) for powder samples placed in magnetic $h_{rf}$ and electric $e_{rf}$ fields for frequency of 9.5 GHz.

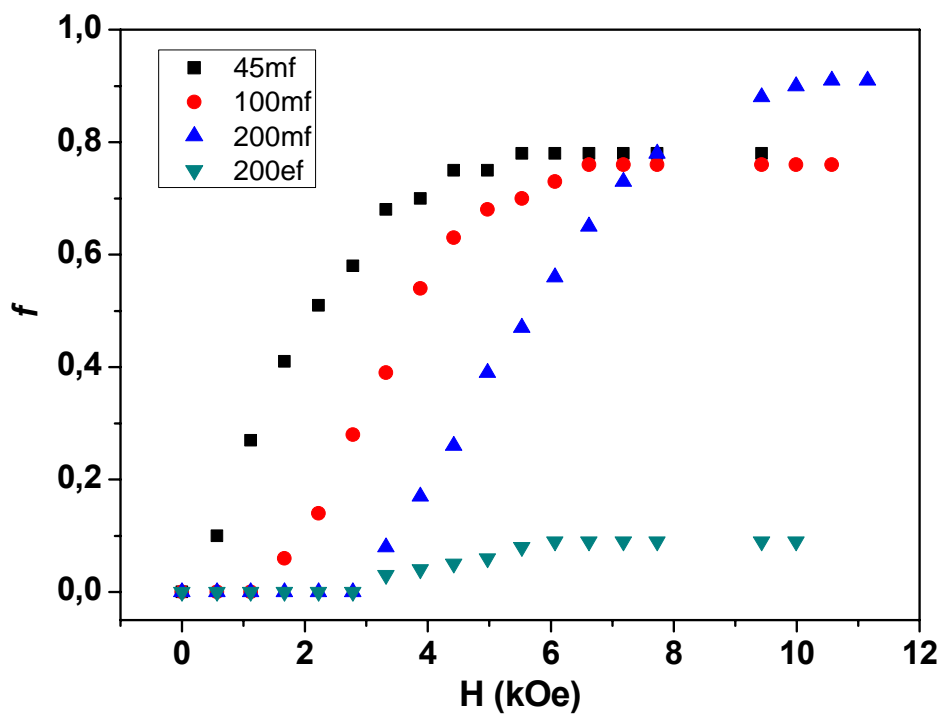

## 5. FINDINGS

**Thin films and multilayers**

A. Permalloy films next to glass substrates

a) In all $Fe_{20}Ni_{80}$ films and multilayers the anisotropy, if any, is uniaxial, with the symmetry axis being close to the film normal.

b) In $Fe_{20}Ni_{80}$ the FMR magnetization $M_{eff}$ = 760 Oe and spin stiffness A = 1.2 x $10^{-6}$ erg/cm while the dc magnetization $M_s$ =800 Oe indicating a uniaxial anisotropy of $K_u \cong$ 2 x $10^5$ erg/cm$^3$.

c) In the FeNi /Cu/FeNi multilayers grown in vacuum $M_{eff}$ in FeNi film on top of Cu is 590 Oe , i.e, significantly lower than that of FeNi. Most likely reason is diffusion of Cu into FeNi during the growth.

d) In the FeNi /Cu/FeNi multilayers if vacuum chamber was "opened" between sputterings, the $M_{eff}$ in FeNi on top of Cu is not diminished. Development of oxide layer inhibits diffusion?

e) Microwave microscope scanning shows that films/layers are reasonably uniform.

**B  Co/GdCo**

a) Generally two types of resonances were observed: one close to resonance of pure GdCo film and another one close to the pure Co film.

b) Clear GdCo resonances in Co(70 Å)/Si or Cu/(Gd-Co)(120 Å))/Si or Cu}$_4$/Co(70 Å) multilayers were observed only for the Si or Cu layers thicker than 7 Å.

c) Resonance field of Co(70 Å)/(Gd-Co)(120 Å) ) }$_4$/Co(70 Å) multilayer is somehow modified being neither at GdCo nor Co field , although closer to Co response. One can think about whole structure response but the magnetization obtained by FMR (close to 800 G) seems too high to be understood without additional study of magnetic properties.

d) Zero field signals were observed in 1-706 and 1-720 samples.

d) Non-magnetic fillers of Si and Cu added to basic Co(70 Å)/(Gd-Co)(120 Å))}$_4$/Co(70 Å) multilayer generally have similar influence on FMR. All low field Co related resonances appear at the fields very close to each other with very weak dependence on the thickness of the non-magnetic layer starting from a few Å thickness. This variation of the resonance field is less pronounced for Cu compared with Si but it is very difficult to say whether observed dependencies indeed reflect the Si and Cu thickness influence. Very thin layers of both Si and Cu seem to change the resonance properties more than layers of 10 Å although it might be that they just prevent the formation of the mixed interfaces as happens in case of pure Co(70 Å)/(Gd-Co)(120 Å))}$_4$/Co(70 Å) structure.

**Dynabeads sytem**

The FMR was measured for a non-aligned sample for 3 different concentrations: resonance fields of 3.2 and 3.4 kOe were obtained with line shape close to Lorentzian. The linewidth is 400 Oe. For the aligned sample the position of the peak was about 3.1 kOe with the same width. $H_{res} \cong$ 3.2 kOe for all cases implies that the g-value is about 2.3. It is likely that there is an effective anisotropy field of about 300 Oe.

**<u>Powders NiCo</u>**

a) SEM shows that powder grains are far from being spherical.

b) All powders have huge zero field absorption as previously reported in micron size grains of Fe + Ni. Application H∥$h_{rf}$ reduces this by 50% at ~ 2-3 kOe. Very little field effect was observed if powder located in $e_{rf}$.

c) FMR observed for first time in nanoparticles. The "line" is wide and overlaps the magneto absorption. Because of the magnetoimpedance effect and the large linewidth measurement of resonance field is far from precise.

d) $H_{res}$ (values were well below those expected for spheres) we attempted to find reasonable explanation treating grains as diskettes with H in-plane, $M_s$ close to the dc value and anisotropy fields (≤ 1 kOe) reducing with increasing grain size. But it seems that uncontrolled local fields are present due to clustering making FMR interpretation problematic.

e) Since the grains are held in glue agglomeration occurs and the samples are not isotropic.

f) Samples aligned in dc field exhibit uniaxial symmetry.

## 5. ACKNOWLEDGMENTS

Galina Kurlyandskaya is deeply grateful to the Ferromagnetic resonance group of the Physics Department of the University of Maryland at College Park for patience, dedication of time and introduction into FMR field during her J1 research Scholarship under program P-1-00793; Valuable assistance of Profs. S.I. Patil and S.E. Lofland, Drs. Sanjay R. Shinde, Sankar Dhar, and A. Ayari is highly acknowledged. All CoNi nanopowders for the measurements were supplied by Prof. M. Vazquez and Mr. C. Luna and therefore the collaboration with Instituto de Ciencia de Materiales (Madrid, Spain) is gratefully acknowledged.